\documentclass[12pt,twoside,a4paper,amsmath,amssymb,apsmy,showkeys,showpacs,nofootinbib]{revtex4-1}
\usepackage[left=2.5cm, right=2.5cm, top=2.5cm, bottom=2.2cm, bindingoffset=0cm]{geometry}
\usepackage{amsfonts}
\usepackage{graphicx,epsfig}

\def\be{\begin{equation}}	\def\ee#1{\label{#1}\end{equation}}
\def\ba{\begin{array}}		\def\ea#1{\label{#1}\end{array}}
\def\bea{\begin{eqnarray}}	\def\eea{\end{eqnarray}}
\def\mc{\mathcal}			\def\pa{\partial}
\def\bi{\bibitem}			\def\ci{\cite}
\def\ra{\rightarrow}		
	\def\I{{\mbox{\scriptsize I}}}
\def\N{{\mbox{\scriptsize N}}}	\def\S{{\mbox{\scriptsize S}}}
\def\GN{\Gamma_\N}			
\def\TOT{{\mbox{\scriptsize TOT}}}	
\def\QCD{{\mbox{\scriptsize QCD}}}

\def\Re{{\rm Re}}	\def\Im{{\rm Im}}

\DeclareGraphicsExtensions{.jpg,.png.,jif,.pdf,.mps,} 

\begin{document}

\title{
Complex Masses of Mesons and Resonances \\In Relativistic Quantum Mechanics}
\author{M. N. Sergeenko}
\email[E-mail: ]{msergeen@usa.com}
\affiliation{
F. Scaryna Gomel state university, Gomel BY-246019, Belarus}

\begin{abstract}
 Relativistic bound state problem in hadron physics is studied. 
 Mesons and their resonance excitations in the framework of 
Relativistic Quantum Mechanics (RQM) are investigated. 
 Two-particle wave equation for the Lorentz scalar QCD inspired 
funnel-type potential with the coordinate dependent strong coupling 
$\alpha_\S(r)$ is derived. 
 The concept of distance dependent particle mass is developed. 
 Two exact asymptotic expressions for the system's squared mass are 
obtained and used to derive the meson interpolating complex-mass formula. 
 Free particle hypothesis for the bound state is developed: quark and 
antiquark move as free particles in of the bound system. 
 Practical applications of the model are given. 
\end{abstract}


\medskip
\medskip
\begin{center}
Reported on XXIV  International  Seminar\\
NONLINEAR PHENOMENA IN COMPLEX SYSTEMS \\
Chaos, Fractals, Phase Transitions, Self-organization \\
May 16--19, 2017, Minsk, Belarus 
\end{center}

\maketitle

\section*{Introduction}\label{intro}
 Most particles listed in the Particle Data Group (PDG) 
tables~\ci{PDG2014} are mesons and their excited states --- {\it resonances}. 
 A thorough understanding of the~physics summarized by the~PDG is related 
to~the~concept of a~resonance. 
 These hadron states are simplest quark-antiquark ($q\bar q$) bound 
systems. 
 The number of known hadrons is constantly increasing with the growing 
energies of accelerators and proposed experiments on LHC~\ci{ATLAS94}. 
 According to the PDG and theoretical predictions, many hadrons are 
still absent from the summary tables. 

 There is no fundamental dynamic theory of hadron resonances at 
the present time. 
 The description of the $q\bar q$ systems and resonances in a way fully 
consistent with all requirements imposed by special relativity and 
within the framework of Quantum Field Theory (QFT) is one of 
the great challenges in theoretical elementary particle physics. 
 Calculations of hadron properties are frequently carried out with 
the~help of phenomenological~\ci{Morp90} and relativistic quark 
models~\ci{Faus11}. 
 One of the~promising among them is the~Regge method in hadron 
physics~\ci{Collin77}. 

 All mesons, baryons  and their resonances in this approach are associated 
with Regge poles which move in the~complex angular momentum $J$~plane. 
 Moving poles are described by the~Regge trajectories, $\alpha(s)$, which 
are the functions of the~invariant squared mass $s=W^2$ (Mandelshtam's 
variable), where $W=E^*$ is the~c.\,m. rest energy (invariant mass of 
two particle system). 
 Hadrons and resonances populate their Regge trajectories which contain all 
the~dynamics of hadron interaction in bound state and scatterin regions. 

 Heavy $Q\bar Q$ mesons (quarkonia) can be considered as nonrelativistic 
(NR) bound systems and well described by the~two-particle Shr\"odinger's 
wave equation. 
 Light $q\bar q$ states are relativistic bound states and require 
another approach. 
 Within the framework of QFT the covariant description of relativistic 
bound states is the Bethe-Salpeter (BS) formalism~\ci{LuchSho16,BethSal08}. 
 The homogeneous BS equation governs all the bound states. 
 However, numerious attempts to apply the BS formalism to relativistic 
bound-state problems give series of difficulties. 
 Its inherent complexity usually prevents to find the exact solutions or 
results in the appearance of excitations in the relative time variable 
of the bound-state constituents (abnormal solutions), which are difficult 
to interpret in the framework of quantum physics~\ci{LuchSho99}. 

 For various practical reasons and applications to both QED and QCD some 
simplified equations, situated along a path of NR reduction, are used. 
 More valuable are methods which provide either exact or approximate 
analytic solutions for various forms of differential equations. 
 They may be remedied in three-dimensional reductions of the BS equation; 
the most well-known of the resulting bound-state equations is the one 
proposed by Salpeter~\ci{Salp52}. 

 In this work we study mesons and their excited states (meson resonances) 
as relativistic two body systems from unified point of view in 
the~framework of the~potential approach. 
 The problem under investigation is related to 1)~two-particle 
relativistic equation of motion and 2)~absence of a~strict definition 
of the potential in relativistic theory. 
 We use the modified funnel type potential in the framework of Relativistic 
Quark Model and treat it as the Lorentz-scalar function of the spatial 
variable~$r$. 
 Using lagrangian relativistic mechanics~\ci{Huang01} we derive the two 
particle classic equation of motion; then, on this basis with the help of 
the correspondence principle we derive the two-body wave equation. 
 We obtain the exact solutions of the equation for two asymptotic 
terms of the potential --- the Coulomb term and linear one. 
 The two asymptotic expressions for the squared mass we use to write 
the complex-mass formula for the bound system. 
 We show that the eigenvalues (masses) for this potential are 
{\it complex} and give several practical applications of the model.

\section{Resonance and its definitions}\label{ResDef}
 There are two well-known definitions of these resonance's parameters, 
both widely used in hadron physics~\ci{BerniCaPe}. 
 One definition, known as the conventional approach, is based on 
the~behavior of the~resonance's {\it phase shift} $\delta(E)$ as 
a~function of the~energy, while the~other, known as 
the~{\it pole approach}, is based on the~pole position of the~resonance 
and includes several approaches~\ci{MorgPenn,Hislop11,Taylor06,MoisPR98}. 
Fundamentals of scattering theory and strict mathematical definition 
of resonances in QM were considered in~\ci{Hislop11,Taylor06,MoisPR98}. 

\begin{figure}[th]
\begin{center}
\includegraphics[scale=0.85]{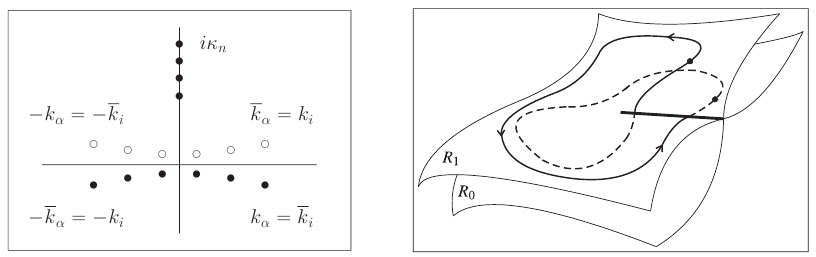}
\caption{{\small Left: Schematic representation of the~poles (disks) 
and the~zeros (circles) of the~transmission amplitude $\mc{A}(k)$ 
in the~complex $k$-plane. Bounded energies correspond to poles located 
on the~positive imaginary axis. 
Right: The~two-sheet Riemann surface $\sqrt{k^2}=\pm k$. 
The~lower edge of the~slit in $R_0$ is joined to the~upper edge of 
the~slit in $R_1$, and the~lower edge of the~slit in $R_1$ is joined to 
the~upper edge of the~slit in $R_0$.}}
\label{fig1:CxPolRiem}
\end{center}\end{figure}

In particle physics resonances arise as unstable intermediate states 
with complex masses~\ci{MoisPR98}. 
 Resonances in QFT are described by the~complex-mass poles of 
the scattering matrix~\ci{Taylor06}. 
 In scattering experiment, the wave function requires different 
boundary condition, that is why the complex energy is 
required~\ci{Taylor06,MoisPR98}. 
 In non-relativistic scattering theory the~complex $k^2=2\mu\mc{E}$ 
variable plane is used. 
 A~Riemann surface $k=\pm\sqrt{2\mu\mc{E}}$ is obtained by replacing 
the~$k^2$-plane with a~surface made up of two sheets $R_0$ and $R_1$, 
each cut along the~positive real axis~\ci{BrowChur} 
(Fig.\,\ref{fig1:CxPolRiem}). 

 The rigorous QM definition of a~resonance requires determining the~pole 
position in the~se\-cond Riemann sheet of the~analytically continued 
partial-wave scattering amplitude in the~complex $\pm k$~variable 
plane~\ci{NieArri}. 
 In relativistic theory, a~Riemann surface $\mc{M}=\pm\sqrt{s}$ 
is obtained by replacing the~$s$-plane with a~surface made up of two 
sheets $R_0$ and $R_1$. 
 Both definitions above are formulated in scattering experiment. 
 On the other hand, there are two basic problems in quantum physics: 
the~scattering problem and bound state problem. 
 Resonances are quasi-stationary states in the $s$-channel 
at $s>0$; this means that one can use another approach to consider 
resonances. 
 Such an approach to bound-state problem is RQM~\ci{GreinerRQM00}. 

 The formulation of RQM differs from non-relativistic quantum mechanics 
by the~replacement of invariance under Galilean transformations with 
invariance under Poincar\`e transformations. 
 The~RQM is also known in the~literature as relativistic Hamiltonian 
dynamics or Poincare-invariant quantum mechanics with direct 
interaction~\ci{Dirac49}. 
 A~definition of resonance in RQM was considered 
in~\ci{MyAHEP13,MyNPCS14}, where mass and width of a~resonance are 
defined from solution of the~eigenvalue problem for the~Cornell 
potential~\ci{EichGMR}, the short-range Coulomb $V_S(r)$ term and 
linear one $V_L(r)$, 
$V(r)=V_S(r)+V_L(r)\equiv-\frac 43\alpha_\S/r +\sigma r$; 
its parameters are directly related to basic physical quantities, 
$\alpha_\S$ and $\sigma$. 

 Operators in ordinary QM are Hermitian and the corresponding 
eigenvalues are real. 
 It is possible to extend the~QM Hamiltonian into the~complex domain 
while still retaining the~fundamental properties of a~quantum theory. 
 This means one can start with the~bound state problem and make 
the~{\it analytic continuation} to the~scattering region. 
 The~problem of particle decay within the~Hamiltonian formalism was 
considered in generalized QM~\ci{SudChiGo}. 
 One of such approaches is complex quantum mechanics~\ci{BendBH}. 
 The~complex-scaled method is the~extension of theorems and principles 
proved in QM for Hermitian operators to non-Hermitian operators. 

 The~Cornell potential is a~special in hadron physics in that sense it 
results in the complex energy and mass eigenvalues. 
 Separate consideration of two asymptotic components of the Cornell 
potential --- $V_S(r)$ and $V_L(r)$ --- results in the complex-masses 
expression for resonances, which in the~center-of-momentum frame (c.m.f.) 
is ($\hbar=c=1$)~\ci{MyAHEP13,MyNPCS14}:
\be 
\mc{M}_\N^2 = 4\left[\left(\sqrt{2\sigma\tilde{N}}
+\frac{i\tilde\alpha m}N \right)^2
+\left(m-i\sqrt{2\tilde\alpha\sigma}\right)^2\right],
\ee{CompE2n}
where $\tilde\alpha_\S=\frac 43\alpha_\S$, $\tilde{N}=N+(k+\frac12)$, 
$N=k+l+1$, $k$ is radial and $l$ is orbit quantum numbers: it has 
the form of the squared energy 
$\mc{M}_\N^2=4\left[(\pi_\N)^2+\mu^2\right]$ of two free relativistic 
particles with the quarks' complex momenta $\pi_\N$ and masses $\mu$. 
 This formula allows to calculate in a~unified way the centered 
masses and total widths of resonances, 
\be 
\mc M_\N^2=\Re\,\mc M_\N^2+i\Im\,\mc{M}_\N^2, \qquad {\rm where}
\ee{E2nReIm}
\be 
\Re\,\mc M_\N^2
= 4\left[2\sigma\tilde N -\left(\frac{\tilde\alpha m}N\right)^2
+m^2-\mu_\I^2\right],\quad 
\Im\,\mc M_\N^2=
8m\mu_\I\left(\frac{\sqrt{\tilde\alpha\tilde N}}N -1\right). 
\ee{ReImE2n}
 The real-part mass in (\ref{ReImE2n}) exactly coincides with the universal 
mass formula obtained independently by another method with the use of 
the two-point Pad\'e approximant~\ci{MyZPhC94} and is very transparent 
physically, as well as the Coulomb potential. 
 It describes equally well the mass spectra of all $q\bar q$ and $Q\bar Q$ 
mesons ranging from the $u\bar d$ ($d\bar d$, $u\bar u$, $s\bar s$) states 
up to the heaviest known $b\bar b$ systems~\ci{MyZPhC94} and 
glueballs~\ci{MyEPJC12,MyEPL10} as well. 
 Besides, it allows one to get the Regge trajectories as analytic functions 
in the whole region from solution of the cubic equation for the angular 
momentum $J(\mc{M}^2)$~\ci{MyZPhC94}; the Regge trajectories including 
the~Pomeron~\ci{MyEPJC12,MyEPL10} are ``saturating'' and appears to be 
successful in many applications~\ci{RossiP03,CLAS_PRL03,CLAS_EPJA05}. 
 
 A~Riemann surface $\mc{M}_\N$ on Fig.\,\ref{fig1:CxPolRiem} can be 
obtained from (\ref{E2nReIm}) by taking the~square root 
$\pm\sqrt{\mc{M}_\N^2}$ replacing the~$\mc{M}_\N^2$-plane with a~surface 
made up of two sheets 
$R_0$ and $R_1$, each cut along the~positive real axis~\ci{BroChur03}. 
 The square root of the complex expression (\ref{E2nReIm}) gives 
\be
\mc{M}_\N=\pm\sqrt{\mc{M}_\N^2}\equiv
\pm\bigl[\Re\,\mc{M}_\N+i\xi\Im\,\mc{M}_\N\bigr], \qquad {\rm where}
\ee{SqCxMn} 
\be
\Re\,\mc{M}_\N=\pm\sqrt{\frac{|\mc{M}_\N^2|
+\Re\,\mc{M}_\N^2}2}={\rm M}_\N, \quad 
\Im\,\mc{M}_\N=\pm\sqrt{\frac{|\mc{M}_\N^2|
-\Re\,\mc{M}_\N^2}2}=-\frac{\GN^\TOT}2. 
\ee{MnGnSq} 
Here $|\mc{M}_\N^2|=\left[(\Re\,\mc{M}_\N^2)^2+
(\Im\,\mc{M}_\N^2)^2\right]^{1/2}$, $\xi={\rm sgn}(\Im\,\mc{M}_\N^2)$. 

 Expressions (\ref{MnGnSq}) give process independent parameters of 
resonances, their centered masses ${\rm M}_\N$ and total 
widths~$\Gamma_\N^\TOT$. 
 The imaginary-part mass in (\ref{MnGnSq}) defines the total width 
$\Gamma_\N^\TOT=-2\Im\,\mc{M}_\N$ of the resonance. 
 The resonance positions are symmetrically located in the Riemann 
$\mc{M}$-surface: if $\mc{M}_p=\Re\mc{M}_p-i\Im\,\mc{M}_p$ is a~pole 
in the~fourth quadrant of the~surface $\pm\sqrt{\mc{M}_\N^2}$, then 
$\mc{M}_p=-\Re\mc{M}_p-i\Im\,\mc{M}_p$ is also a~pole, but in the~third 
quadrant (antiparticle)~\ci{MoisPR98}.

\section{ Relativistic two-body problem}\label{Rel2Body}
Considered above quarkonia are simplest among mesons. 
A more complicated case represent mixed $Q\bar q$ bound states in which 
quark masses are different. 
Solution of the~relativistic two-body (R2B) problem in a~way fully 
consistent with all requirements imposed by special relativity and within 
the~framework of QFT is one of the~great challenges in theoretical 
elementary particle physics~\ci{CagnAll94}. 
Comprehensive description of $Q\bar q$ bound states is reduced 
to relativistic bound state problem. 

There have been proposed several wave equations for the describtion of 
bound states within relativistic quantum theory such as the~Klein-Gordon, 
the~Dirac, quasipotential, etc. 
Relativistic description of two-body systems is usually based on 
the~four-dimensional covariant BS equation~\ci{BethSal08}. 
This equation governs all the~bound states and is appropriate framework 
for the~description of the~R2B problem within QFT.  
However, attempts to apply the~BS formalism to the~R2B problem give 
series of difficulties, the interaction kernel entering in this 
equation is not derivable from the~first principles. 

All these difficulties are the sources of the numerous attempts 
to reformulate the~R2B problem of QFT~\ci{LuchSho16,Faus11}; 
there exist various reductions of the~BS equation (see~\ci{LuchSho16}). 
Many authors have developed noncovariant instantaneous truncations of 
the~BS equation~\ci{LuchSho99,LuchSho16}, but a~better known is 
the~Salpeter work~\ci{Salpet52}. 
After applying simplifying assumptions and approximations to the~BS 
equation, one comes to the~spinless Salpeter equaton, which in the~c.m.f. 
is~\ci{Salpet52,LuchSho16} 
\be 
\left[\sqrt{\mathbf{\hat p}^2+m_1^2}
+\sqrt{\mathbf{\hat p}^2+m_2^2}+V(\mathbf{r})\right]\psi=\sf{M}\psi;
\ee{SalpEq}
it is the conceptually simplest bound-state wave equation incorporating 
to some extent relativistic effects, where the~potential $V(\mathbf{r})$ 
arises as the~Fourier transform of the~BS kernel $\mc{K}(\mathbf{q})$. 
The equation (\ref{SalpEq}) is sometimes denoted semirelativistic 
since is not a~covariant formulation, however, even this simplest 
two-particle wave equation leads to several 
difficulties~\ci{LuchSho99,LuchSho16}. 

The square roots of the operator in (\ref{SalpEq}) cannot be used as 
they stand; they would have to be expanded in a power series before 
the squared momentum operator, $\mathbf{\hat p}^2$, raised to a~power 
in each term, could act on $\psi(r)$. 
As a result of the power series, the space and time derivatives are 
completely asymmetric: infinite-order in space derivatives but only 
first order in the time derivative, which is inelegant and unwieldy. 
 The next problem is the noninvariance of the energy operator 
in~(\ref{SalpEq}), equated to the square root which is also not invariant. 
 A~more severe problem is that it can be shown to be nonlocal and 
can even violate causality~\ci{LuchSho16}. 

 The~spinless Salpeter equaton (\ref{SalpEq}) can be obtained by another 
way without any approximations in the framework of RQM from the following 
simple and obvious consideration. 
 The underlying classic equation of (\ref{SalpEq}) is 
\be 
\sqrt{\mathbf{p}^2+m_1^2}+\sqrt{\mathbf{p}^2+m_2^2}+V(r)=\sf{M}.
\ee{Class2B}
 This is the total energy of two interacting particles in the~c.m.f. 
and, at the same time, the mass of the system which can also be found 
from the following simple consideration. 

Equation (\ref{Class2B}) is the zeroth component of the four-vector 
$\mc{P}^\mu=p_1^\mu+p_2^\mu+W^\mu(q_1,\,q_2)$. 
Two particles with the momenta $p_1$, $p_2$ and the interaction field 
$W^\mu(q_1,\,q_2)$ (potential) together represent a~closed system. 
The four-vector $\mc{P}^\mu$ describes free motion of the~bound system 
and can be separated into the two equations, i.\,e., 
$E=\left[\mathbf{p}_1^2+m_1^2\right]^{1/2}
+\left[\mathbf{p}_2^2+m_2^2\right]^{1/2}
+W^0(q_1,\,q_2)=\rm{const}$ and 
$\mathbf{P}=\mathbf{p}_1+\mathbf{p}_2+\mathbf{W}(q_1,\,q_2)=\rm{const}$, 
describing the~energy and momentum conservation laws. 
 Because the~energy $E$ of the~system is the~constant of motion, 
the~corresponding Hamiltonian can not depend on time explicitly, 
the~potential $W(q_1,\,q_2)$ should not depend on time, but depends on 
the~relative coordinat $\mathbf{r}=\mathbf{r}_1-\mathbf{r}_2$ of particles, 
i.\,e., $W^0=S(\mathbf{r})$~\ci{LuchSho16,LuchSho99}. 
 This means that the~system's relative time is zero, $\tau=0$, and 
the~interaction is instantoneous, the total energy $E$ in the~c.m.f. is 
the~mass \textsf{M} of the system. 

 The potential in (\ref{Class2B}) is the zeroth component of 
the~Lorentz-vector $W^\mu(q_1,\,q_2)$. 
 In general, there are two different relativistic versions: the~potential 
is considered either as the~zero component of a~four-vector, or as 
a~Lorentz-scalar~\ci{SahuAll89}; its nature is a~serious problem of 
relativistic potential models~\ci{Sucher95}. 
 The~relativistic correction for the~case of the~Lorentz-vector 
potential is different from that for the~case of the~Lorentz-scalar 
potential, the~radial $S$-wave function $R(r)$ for the~hydrogen atom 
diverges as $r\ra 0$. 
 This problem is very important in hadron physics where, for 
the~vector-like confining potential, there are no normalizable 
solutions~\ci{Sucher95,SemayCeu93}. 
 There are normalizable solutions for scalar-like potentials, but not 
for vector-like. 
 This issue was investigated in ~\ci{MyZPhC94,Huang01};  
the~effective interaction has to be Lorentz-scalar in order 
to confine quarks and gluons. 

 The relativistic total energy $\epsilon_i(\mathbf{p}_i)$ of a~particle 
in (\ref{Class2B}) given by 
$\epsilon_i^2(\mathbf{p}_i)=\mathbf{p}_i^2+m_i^2$ can be represented as 
sum of the kinetic energy $\tau_i(\mathbf{p})$ and the rest mass $m_i$, 
i.\,e., $\epsilon_i(\mathbf{p})=\tau_i(\mathbf{p})+m_i$. 
 Then the~energy-mass $\sf{M}$ in (\ref{Class2B}) can be written as 
$\mathsf{M}=[\tau_1(\mathbf{p})+m_1]+[\tau_2(\mathbf{p})+m_2]+V(r)$, 
the potential $V(r)$ can be considered as scalar-like $\mathsf{S}(r)$ 
and shared among the two masses $m_1$ and $m_2$ as 
$\mathsf{m}_i(r)=m_i+\frac 12\mathsf{S}(r)$~\ci{MesResX17}. 
 Therefore, the system's total energy $E$ (invariant system mass 
\textsf{M}) can be written in the form 
\be 
\mathsf{M} = \sqrt{\mathbf{p}^2+\mathsf{m}_1^2(r)}
+\sqrt{\mathbf{p}^2+\mathsf{m}_2^2(r)}.
\ee{ClaMasr}
 The functions $\mathsf{m}_i(r)$ in (\ref{ClaMasr}) are treated as the 
distance dependent particle masses and (\ref{ClaMasr}) can be transformed 
to the~squared relative momentum, 
$\mathbf{p}^2=\left[s-m_-^2\right]\left[s-(m_+ +\mathsf{S})^2\right]/4s$, 
where $s=\sf{M}^2$. 
 This expression with the~help of the~fundamental correspondence principle 
(according to which physical quantities are replaced by operators acting 
onto the~wave function) gives the~two-particle spinless wave 
equation~($\hbar=c=1$), 
\be 
\Bigl\{\vec{\nabla}^2+K(s)\left[\mathsf{M}^2
-\left(m_+ +\mathsf{S}\right)^2\right]\Bigr\}\psi(\vec r)=0,
\ee{Rel2Eq}
where $K(s)=(s-m_-^2)/4s$, $m_+=m_1+m_2$, $m_-=m_1-m_2$. 

 The~Cornell potential we consider here is fixed by the~two free 
parameters, $\alpha_\S$ and $\sigma$. 
 However, the~strong coupling $\alpha_\S$ in QCD is a~function 
$\alpha_\S(Q^2)$ of virtuality $Q^2$ or $\alpha_\S(r)$ in configuration 
space. 
 The potential can be modified by introducing the $\alpha_\S(r)$-dependence, 
which is unknown. 
 A~possible modification of $\alpha_\S(r)$ was introduced in~\ci{MyEPJC12}, 
\be
V_\QCD(r) = -\frac 43\frac{\alpha_\S(r)}r +\sigma r,\quad 
\alpha_\S(r)=\frac 1{b_0\ln[1/(\Lambda r)^2+(2\mu_g/\Lambda)^2]},
\ee{VmodCor}
where $b_0=(33-2n_f)/12\pi$, $n_f$ is number of flavors, 
$\mu_g=\mu(Q^2)$ --- gluon mass at $Q^2=0$, $\Lambda$ is the~QCD scale 
parameter. 
The~running coupling $\alpha_\S(r)$ in (\ref{VmodCor}) is frozen at  
$r\ra\infty$, $\alpha_\infty=\frac 12[b_0\ln(2\mu_g/\Lambda)]^{-1}$, and 
is in agreement with the~asymptotic freedom properties 
[$\alpha_\S(r\ra 0)\ra 0$].

\section{ Solution of the~wave equation}
\label{SolQCEq}
 It is a~problem to find the~analytic solution of known equations, as well 
as (\ref{Rel2Eq}), for the~potential~(\ref{VmodCor}). 
 Instead, we solve the~quasiclassical (QC) wave equation~\ci{MyMPLA97,MyPRA96}. 
 Derivation of the~QC equation is reduced to replacement of 
the~operator $\vec{\nabla}^2$ by the~canonical operator 
$\Delta^c$~\ci{MyPRA96} without the~first derivatives, acting onto 
the~state function $\Psi(\vec r)$. 
 Therefore, instesd of (\ref{Rel2Eq}) we solve the~QC equation for 
the~potential~(\ref{VmodCor}), 
\be 
\Biggl\{\frac{\pa^2}{\pa r^2}+\frac 1{r^2}\frac{\pa^2}{\pa\theta^2}
+\frac 1{r^2\sin^2\theta}\frac{\pa^2}{\pa\varphi^2}
+K(s)\biggl[\mathsf{M}^2-\left(m_+ +V_\QCD\right)^2\biggr]
\Biggr\}\Psi(\vec r)=0,
\ee{Rel2Equa}
which is separated that gives the radial,
\be 
\Biggl\{\frac{d^2}{dr^2}+\frac{s-m_-^2}{4s}
\biggl[s-\left(m_+ -\frac 43\frac{\alpha_\S(r)}r+\sigma r\right)^2\biggr] 
-\frac{\textrm{M}_l^2}{r^2}\Biggr\}\textrm{R}(r)=0,
\ee{QCrad}
and the~angular equations. 
Solution of the~last was obtained in~\ci{MyPRA96} by the~QC method in 
the~complex plane, that gives $\textrm{M}_l=(l+\frac 12)\hbar$, 
for the~angular momentum eigenvalues. 
These angular eigenmomenta are universal for all spherically symmetric 
potentials in relativistic and NR cases. 

The~problem (\ref{QCrad}) has four turning points and cannot be 
solved analytically by standard methods. 
We consider the~problem separately by the~QC method for the~short-range 
Coulomb term and the long-range linear term. 
The~QC wave equation (\ref{QCrad}) for the~Coulomb term has two turning 
points and the phase-space integral is 
\be
I=\oint_C\sqrt{\frac{s-m_-^2}{4s}\biggl[s-\left(m_+ 
-\frac 43\frac{\alpha_\S(r)}r\right)^2\biggr] 
-\frac{(l+\frac 12)^2}{r^2}}\,dr = 2\pi\left(k+\frac 12\right).
\ee{FasIcl}
The~phase-space integral (\ref{FasIcl}) is found in the~complex plane 
with the~use of the~method of stereographic projection~\ci{MyAHEP13} 
that gives 
\be
\mathsf{M}_\N^2 
=\left(\sqrt{\epsilon_\N^2}\pm \sqrt{(\epsilon_\N^2)^*}\right)^2 
\equiv 4\left[\Re\{\epsilon_\N^2\}\pm i\Im\{\epsilon_\N^2\} \right],
\ee{W2Coul}
where $\epsilon_\N^2
=\frac 14 m_+^2\left(1-v_\N^2\right) +\frac i2 m_+ m_-v_\N$, 
$v_\N=\frac 23\alpha_\infty/N$, $N=k+l+1$. 

 Large distances in hadron physics are related to the~problem of 
confinement. 
 The~problem has four turning points, i.\,e., two cuts between these 
points. 
 The~phase-space integral (\ref{FasIcl}) is found by the~same method of 
stereographic projection as~above that results in the~cubic equation: 
$s^3 + a_1s^2 + a_2s + a_3 = 0$, where 
$a_1=16\tilde\alpha_\infty\sigma-m_-^2$, 
$a_2=64\sigma^2\left(\tilde\alpha_\infty^2-\tilde N^2
-\tilde\alpha_\infty m_-^2/4\sigma\right)$, 
$a_3=-(8\tilde\alpha_\infty\sigma m_-)^2$, $\tilde N=N+k+\frac 12$, 
$\tilde\alpha_\infty=\frac 43\alpha_\infty$. 
 The~first root $s_1(N)$ of this equation gives the~physical 
solution (complex eigenmasses), $\mathsf{M}_1^2(N)=s_1(N)$, for 
the~squared invariant mass. 

Two exact asymptotic solutions, i.\,e., (\ref{W2Coul}) and the~first root 
of the~cubic equation above, are used to~derive the~resonance's mass 
formula. 
The~interpolation procedure for these two solutions is used~\ci{MyZPhC94} 
to~derive the~resonance's mass formula:
\be
\mathsf{M}_\N^2 
=\left(m_1+m_2\right)^2 \left(1-v_\N^2\right)\pm 2i(m_1^2-m_2^2)v_\N 
+\Re\{\mathsf{M}_1^2(N)\}.
\ee{W2int}
 The real part of the square root of (\ref{W2int}) defines the~centered 
mass and its imaginary part defines the~total widths, 
$\Gamma_\N^\TOT=-2\,\Im\{\mathsf{M}_\N\}$, of 
the~resonance~\ci{MyAHEP13,MyNPCS14}. 

To demonstrate its efficiency we calculate the leading-state masses of 
the $\rho$ and $D^*$ meson resonances (see tables, where masses are in MeV). 

\begin{table}[ht]
\begin{center}
\caption{The masses of the $\rho^\pm(u\bar d)$-meson resonances}
\label{rho_mes}
\begin{tabular}{lllll}
\hline\noalign{\smallskip}
\ \ Meson &~~~$J^{PC}$ &~~~$\ \ E_n^{ex}$ &~~~$\ \ E_n^{th}$&
~~~Parameters in (\ref{W2int})\\
\noalign{\smallskip}\hline\hline\noalign{\smallskip}
\ \ \ $\rho\ (1S)$&~~~$1^{--}$&~~~$\ \ 776$&~~~$\ \ 776$&~~~~~$\Lambda=500$ MeV\\
\ \ \ $a_2(1P)$&~~~$2^{++}$&~~~$\ 1318$&~~~$\ 1314$&~~~~~$\mu_g=416$\,MeV\\ 
\ \ \ $\rho_3(1D)$&~~~$3^{--}$&~~~$\ 1689$&~~~$\ 1689$&~~~~~$\sigma=0.139$\,GeV$^2$\\ 
\ \ \ $a_4(1F)$&~~~$4^{++}$&~~~$\ 1996$&~~~$\ 1993$&~~~~~$m_d=276$\,MeV\\ 
\ \ \ $\rho\ (1G)$&~~~$5^{--}$&~~~$ ~ $&~~~$\ 2255$&~~~~~$m_u=129$\,MeV\\
\ \ \ $\rho\ (2S)$&~~~$1^{--}$&~~~$\ 1717$&~~~$\ 1682$&~\\
\ \ \ $\rho\ (2P)$&~~~$2^{++}$&~~~$ ~ $&~~~$\  1990$&~\\
\ \ \ $\rho\ (2D)$&~~~$3^{--}$&~~~$ ~ $&~~~$\ 2254$&~\\
\noalign{\smallskip}\hline
\end{tabular}
\caption{The masses of the $D^{*\pm}(c\bar d)$-meson resonances}
\label{Dp_mes}
\begin{tabular}{lllll}
\hline\noalign{\smallskip}
\ \ Meson &~~~$J^{PC}$ &~~~$\ \ E_n^{ex}$ &~~~$\ \ E_n^{th}$&
~~~Parameters in (\ref{W2int})\\
\noalign{\smallskip}\hline\hline\noalign{\smallskip}
\ \ \ $D^*(1S)$&~~~$1^{--}$&~~~$\ 2010$&~~~$\ 2010$&~~~~~$\Lambda=446$\,MeV\\
\ \ \ $D_2^*(1P)$&~~~$2^{++}$&~~~$\ 2460$&~~~$\ 2464$&~~~~~$m_g=416$\,MeV\\ 
\ \ \ $D_3^*(1D)$&~~~$3^{--}$&~~~$\ ~ $&~~~$\ 2845$&~~~~~$\sigma=0.249$\,GeV$^2$\\ 
\ \ \ $D_4^*(1F)$&~~~$4^{++}$&~~~$\ ~ $&~~~$\ 3178$&~~~~~$m_c=1163$\,MeV\\
\ \ \ $D_5^*(1G)$&~~~$5^{--}$&~~~$\ ~ $&~~~$\ 3478$&~~~~~$m_d=271$\,MeV\\
\ \ \ $D^*(2S)$&~~~$1^{--}$&~~~$\ 1820$&~~~$\ 2821$&~\\
\ \ \ $D^*(2P)$&~~~$2^{++}$&~~~$\ 2011$&~~~$\ 3166$&~\\
\ \ \ $D^*(2D)$&~~~$3^{--}$&~~~$\ ~ $&~~~$\ 3471$&~\\
\noalign{\smallskip}\hline
\end{tabular}
\end{center}
\end{table}

The~free fit to the~data show a~good agreement for the~light and heavy 
$Q\bar q$ meson resonances. 
Note, that the gluon mass in the~independent fitting is the~same, 
$m_g=416$\,MeV. 
Besides, it is the~same for glueballs~\ci{MyEPJC12}. 
The~$d$ quark effective mass is also practically the same, i.\,e., 
$m_d\simeq 273$\,MeV, for the light and heavy resonances. 

\section*{Conclusion} 
\label{Conclu}
The constituent quark picture could be questioned since potential models 
have serious difficulties because the potential is non-relativistic concept. 
However, in spite of non-relativistic phenomenological nature, 
the~potential approach is used with success to describe mesons as~bound 
states of quarks. 

We have modeled meson resonances to be the~quasi-stationary states of two 
quarks interacting by the~QCD-inspired funnel-type potential with 
the~coordinate dependent strong coupling, $\alpha_\S(r)$. 
Using the complex analysis, we have derived the meson complex-mass 
formula~(\ref{W2int}), in which the real and imaginary parts are exact 
expressions. 
This approach allows to simultaneously describe in the unified way 
the centered masses and total widths of resonances. 
We have shown here the~results only for unflavored and charmed meson 
resonances, however, we have obtained a~good description for strange and 
beauty mesons as well~\ci{MesResX17}. 


\end{document}